# First Things First

If Software Engineering is the Solution, then What is the Problem?

Jesús Zavala Ruiz
Universidad Autónoma Metropolitana - Iztapalapa
Mexico City, Mexico.
e-mail: jzr@xanum.uam.mx, jzavalar@acm.org

*Abstract*— Software engineering (SE) undergoes an ontological crisis and it lacks of a theory. Why? Among other reasons, because always it succumbed to the pragmatism demanded by the commercial and political interests and abandoned any intention to become a science instead of a professional discipline. For beginning a discussion for define a theory of software, first, is required to know what software is.

## I. Introduction

Finally, we are in an historical moment in which is possible to discuss, philosophical and politically, about the essence of *such thing* called software engineering (SE), at least, not as open as now. SE is under a severe crisis but it is not the so called "software crisis" which emerged in the 1960s. Now, it's a disciplinary or ontological crisis. The state of the art of SE, without a well rooted theory, confirms our spectacular failure. So, we need to define the *essence of a science of software* by shifting us to a new paradigm but abandoning the old and dominant one. There isn't another way. If we don't do it, we'll not create a theory of software, solid and useful [1]

## II. Methodology

First, we need to recognize that, in computer sciences (CSs) (in plural); there isn't uniformity in concepts, approaches, methodologies, techniques and philosophies. *CS is a wide umbrella* where diversity has flourished, pushing to innovation, technology and science to frontiers never imagined, sometimes *breaking the established rules*. It's possible only when, in action, a *collage* of ontologies, epistemologies and methodologies *is tolerant* to the others. This way emerges as a collage of eclectic, anarchical and opportunistic nature [2]. This has occurred, sometimes, *without a previous theory* but it has flourished after experimenting. SE as another computer branch responds to this pattern, too. So, I used a methodology based on these ideas but as suggested by E. Morin's for his complexity theory, but in a sociological approach instead of a technical one. For begin, I proposed a multilevel framework of the software of ontologies.

## III. The history of SE matters

In an approach based on the Science & Technology Studies is clear that SE was founded when the interests of the first huge customer of software, the U.S. Department of Defense (DoD) and its contractors, could not defeat the interests of the ACM (the main professional association of that time) in an industry of software, academically strong, but DoD was looking for a silver bullet for its managerial crisis and to control to its huge software projects and to its contractors. So, the software crisis term was coined. In consequence, the First SE Conference was celebrated in Garmisch, in 1968, strangely enough, without the invitation to the ACM. After this, would was evident an intellectual and political hostility towards the ACM. The political intention to replace "computer sciences" with "software engineering" would was so evident [3]. An antagonism arose between the ACM and those whom supported SE and created the IEEE-CS. Today, that historical disagreement has not gone away and it continues a subtle intention for "filing for divorce" between them. In any case, SE was invented in Garmisch, at least in a rhetorical sense as the technical and management discipline that would solve the software crisis. The *silver bullet* was discovered, finally.

The main consequence of this history of computing is that SE has not played the major role that it would have wanted. CS did not hesitate and now, more than fifty years after of its foundation, is widely recognized as a science and beyond, it has influenced those others. However, SE succumbed to the outside interests and now it navigates to the drift, without such a progress. Now there is no longer reason to such rivalry. The SE did fail and the closest thing to a theory of software is: the SWEBoK [4] supported by the IEEE-CS (made to the image and likeness of the PMBoK Guide), the models for management and organization for software contractors of the CMM-I by Software Engineering Institute (SEI) and recently, the SEMAT Initiative [1]. So, the question is: *how to develop a theory of software?*

TABLE I. Ontological levels of software

| Level | Dimensions | | |
|---|---|---|---|
| | *Ontology* | *Epistemology* | *Methodology* |
| 4 | Software Factory (Organization) | Sociology of Organization Theory | Business Organization |
| 3 | Software Project (Project) | Project Management Theory | Project Management |
| 2 | Software Dev. (Production Process) | Work Organizing Theory | Process Engineering |
| 1 | Computing & Software System (Software) | ¿Software Science & Software Engineering? | Software Development |

Source: Own elaboration from [1]

## IV. THE ONTOLOGY OF SOFTWARE: WHAT IS SOFTWARE?

I propose this way: First, define what software is (ontology). Second, discover how we can know it in a scientific manner (epistemology and methodology). Third, establish how we can to practice it (praxeology). Finally, how we could behave in a fair and proper manner (axiology) [1]. We have to try this way because the inverse one has failed in our history. So first things first: the ontology.

### A. First ontological level: Computing & software system

This level focuses on the software as computing and system. This is the technical dimension closer to its definition [5], formal but very narrow in practice. CS has worked on this and there are some theories and techniques developed on low and high level. By contrast, SE has tried to focus, mistakenly, only at high abstraction levels. For example, requirements engineering gives an idea about the huge complexity of software. The 2012 revision of the ACM Computing Classification System (CCS) gives a best panorama [6]. OOA&D, UML among others belong to this level.

### B. Second ontological level: The development process

This second level focuses on the *software as development process* that, in essence, is an organizing process of the individual and team work. Due to its symbolic and social nature, this ontology is a concern of organization theories and management science. However, it is a main focus in SE because it has defined as a technical and managerial challenge. It has tried to discover the laws of software development without a full success.

### C. Third ontological level: The software project

Here, the focus is *the software as project*. It's a temporary organizational form of resources and people. It's a unique organizational and managerial process in a social and institutional milieu. It's the central concern of project management. Recently, has criticized itself the relevance of its traditional theory based on the "PMBoK Guide" and it's looking for a most suitable solution [7]. SE has had concerns in this level too, as evidenced by the RUP among others.

### D. Fourth ontological level: The software factory

This level focuses on *the software as factory* [8], as an organization while operating in a complex but specific environment defined by its socio-technical, institutional and cultural dimensions. The organizational sociology and entrepreneurship theory address this level, but SE too, for example, the CMM-I developed by SEI.

## V. DISCUSSION

As we can see, the software is a material and symbolic complex object which transcends its material dimension in a knowledge economy. In SE, there is two complementary streams: (1) the official software engineering (traditional), guided by the interests of the large multinational companies, large customers and manufacturers, governments and its agencies (like DoD), in general by the market, and (2) the alternate free-libre and open source software engineering more guided by freedom and knowledge as social commons than by profits. Apparently, the unique difference between the two SE is the underlying paradigms of management. So, both have succeeded and failed *without an explicit theory of software* or such theory doesn't exist.

As we can see, SE has tried to address all these ontologies but in a disordered mixture, so we ended quite confused. This framework does make possible to manage, progressively, ontologies located in very distinct levels of complexity making more explicit its relationships. So, it's possible to visualize that only the first ontological level is appropriate to being concerned of SE because the other three upper levels are concerns of management science and organization science, mainly. If we don't accept this then we have to recognize that SE a management field, but not engineering.

## VI. CONCLUSIONS

The *science of software*, whatever it may be named, should concern only of the first ontological level. The two currents of SE are the two faces of the same coin, and then it's indispensable to invite to share us, in SEMAT, the vision and experience of free-libre and open source software movement. The SE and the CS should reconcile themself in an act of intellectual and political humility and accept that the SE or the science of software is the missing science. It seems that ACM already has given the first step to because it has integrated in the "Software and its Engineering" section in CCS, *all subjects related to software*. Finally, because SE was caged into its own "iron cage", it's very difficult to hope a paradigmatic shift inside it, so I propose to do it by *creating the science of software as the theoretical field of software*. So, we need to create a new language which incorporates other vocabularies and constructs a multidisciplinary umbrella, as our challenge demands it. If we don't succeed doing this synthesis, coming years we will continue questioning: *if software engineering is the solution, then what's the problem?*